\documentstyle[aps,preprint]{revtex}
\tighten
\newcommand{\st}{\stackrel}
\newcommand{\lh}{\leftrightarrow}
\newcommand{\pp}{\partial}

\newcommand{\p}{\perp}
\newcommand{\be}{\begin{eqnarray}}
\newcommand{\e}{\end{eqnarray}}
\begin{document}
\tighten
\title{\bf Transverse Spin in QCD and Transverse Polarized Deep
Inelastic Scattering}
\author{{A. Harindranath}, 
{ Asmita Mukherjee} \\
{\it Saha Institute of Nuclear Physics, 1/AF, Bidhan Nagar, 
	Calcutta 700064 India} \\
{Raghunath Ratabole} \\
 {\it Centre for Theoretical Studies, Indian Institute of Science \\
     Bangalore 560012 India} \\}
\date{November 4, 1999}
\maketitle
\begin{abstract}
We address the long standing problem of the construction of relativistic spin
operators for a composite system in QCD.  Exploiting the kinematical boost 
symmetry in light front theory, we show that transverse spin operators 
for massless  particles can be introduced in an  arbitrary reference frame, 
in analogy with  those for massive particles. In light front QCD, the complete
set of  transverse spin operators are identified for the first time, which are
responsible for the helicity flip of the nucleon. We   
establish the direct connection between transverse spin in light front QCD and 
transverse polarized deep inelastic scattering. We discuss the theoretical
and phenomenological implications of our results.                          
\end{abstract}
\vskip .2in
The complexity of spin of a composite system in the equal time quantization
of relativistic quantum field 
theory is well-known\cite{alfaro,bou}. The closest one can get to covariant spin
operators are the Pauli-Lubanski operators. In the equal time quantization,
they readily qualify for spin operators {\it only} in the rest frame of 
the particle.
How to construct the spin operators for a composite system in an arbitrary
reference frame is a nontrivial problem. 
The complexities arise
from the facts that for a moving composite object, {\it Pauli-Lubanski
operators are necessarily interaction dependent} and, further, it is quite
difficult to separate the center of mass and internal variables\cite{osb}. 
Due to these difficulties, the canonical structure of spin operators of a
composite system in a moving frame in gauge field theory has never been 
studied in equal time quantization.

In the light front formulation of quantum field theory, in addition to the
Hamiltonian $P^-$, the two rotation operators $F^i$ are interaction
dependent. Moreover, even for massive particles, 
together with the helicity operator ${\cal J}^3$ they
obey $E_2$-like algebra, not the angular momentum algebra 
appropriate for spin operators of a massive particle. In this case, how to 
define the appropriate spin operators is known\cite{ls78}.  
Most of the studies of the transverse spin operators, so far, are
restricted to free field theory. Even in this case the operators have a
complicated structure, however, one can 
write these operators as a
sum of orbital and spin parts via a unitary
transformation, the famous Melosh transformation\cite{melosh}. 
In interacting
theory, presumably this can be achieved order by order\cite{bp} in a suitable
expansion parameter 
which is justifiable only in a weakly  coupled theory.

Very little is
known\cite{review} regarding the field theoretic aspects of the
interaction dependent spin operators, knowledge about 
which is mandatory for issues concerning Lorentz invariance in
light front theory. In this work we show that, in spite of the complexities,
light front field theory
offers a unique opportunity to address the issue of relativistic spin
operators in an arbitrary reference frame since boost is kinematical in this
formulation.

 From the phenomenological point of view, the issue of transverse spin has 
become very important in high energy physics thanks to recent 
experimental advances\cite{expt}. Since transverse spin for a free massless
gluon is identically zero, transverse spin measurements for gluonic
observables directly probe the long distance, nonperturbative features of
QCD. Analogous to longitudinally polarized scattering, where 
quark helicity carries roughly only 25 \% of the proton helicity, 
one may ask what is the
situation in transversely polarized scattering. 
In particular can one relate the operators appearing in the transverse spin
to the integrals of structure functions appearing in transverse polarized
scattering? 

In terms of the gauge invariant, symmetric energy momentum tensor
$\Theta^{\mu \nu}$ the four-vector $P^\mu$ and the
generalized angular momentum 
tensor $M^{\mu \nu}$ are given by
\begin{eqnarray}
P^\mu &&= {1 \over 2} \int dx^- d^2 x^\perp \Theta^{+ \mu},  \\
M^{\mu \nu} && = {1 \over 2} \int dx^- d^2 x^\perp \left [ x^\mu \Theta^{+
\nu} - x^\nu \Theta^{+ \mu} \right ]. 
\end{eqnarray}
The boost operators are $ M^{+-} = 2 K^3$ and $M^{+i}=E^i$. The rotation
operators are $ M^{12}=J^3$ and $ M^{-i} = F^i$. The Hamiltonian $P^-$ and
the transverse rotation operators $F^i$ are dynamical (depend on the
interaction) while the other seven operators are kinematical. The 
rotation operators obey the algebra of two dimensional Euclidean space,
namely,
\begin{eqnarray}
[F^1,F^2]=0, ~ [J^3,F^i] = i \epsilon^{ij} F^j \label{eq1}
\end{eqnarray}
where $\epsilon^{ij}$ is the two-dimensional antisymmetric tensor. Thus $F^i$
do not qualify as angular momentum operators. Moreover, since they do not
commute with $P^\mu$, they do not qualify as spin operators.

Consider the Pauli-Lubanski spin operators 
\begin{eqnarray}
W^\mu = - { 1 \over 2} \epsilon^{\mu \nu \rho \sigma} M_{\nu \rho}
P_\sigma 
\end{eqnarray}
with $ \epsilon^{+-12} = -2$.
For a massive particle, the transverse spin operators\cite{ls78} ${\cal J}^i$ in 
light front theory are given in terms of Poincare generators by
\begin{eqnarray}
M{\cal J}^i &&= W^i - P^i {\cal J}^3 ~~~(i=1,2) \\
&&= \epsilon^{ij}\Big(   { 1 \over 2} F^j P^+ - {1 \over 2} 
E^j P^- + K^3 P^j  \Big )- P^i {\cal J}^3.    
\end{eqnarray}
The interaction dependence of ${\cal J}^i$ arises from $F^i$ which depends
on both center of mass and internal variables. The rest of the terms in
${\cal J}^i$ serves to remove the center of mass motion effects from $F^i$. 
The helicity operator
\begin{eqnarray}
{\cal J}^3 &&= {W^+ \over P^+} = J^3 + { 1 \over P^+}(E^1P^2 - E^2 P^1)
\label{j3} 
\end{eqnarray}  
which is interaction independent. The last two terms in ${\cal J}^3$ removes
the center of mass motion effects from $J^3$. 
We also have
\begin{eqnarray}
W^- && = F^2 P^1 - F^1 P^2 - J^3 P^- .   
\end{eqnarray}
The operators ${\cal J}^i$  ($i=1,2,3$) obey the angular momentum 
commutation relations 
\begin{eqnarray}
\left [ {\cal J}^i, {\cal J}^j \right ] = i \epsilon^{ijk} {\cal J}^k
  .
\end{eqnarray}
For a single fermion of mass $m$, momenta $(k^+,k^\perp)$ and helicity
$\lambda$, we get $ {\cal J}^3 \mid k \lambda \rangle = { \lambda \over 2 }
\mid k \lambda \rangle$, $ \lambda = \pm 1$;
${\cal J}^i \mid k \lambda \rangle = { 1 \over 2}
\sum_{\lambda'}\sigma^i_{\lambda' \lambda} \mid k \lambda'
\rangle,$ $ i= 1,2$.

In the case of massless particle, for the light-like vector $ p^\mu$, 
usually the collinear choice is
made\cite{Tung,Weinberg}, namely, $p^+ \neq 0$, $
p^\perp=0$. For calculations with composite states we
would like to have results for a light-like particle with arbitrary
transverse momenta. Let us try a light like momentum $ P^\mu$ with $ P^\perp
\neq 0$, but $ P^- = {(P^\perp)^2 \over P^+}$ so that $P^2 = 0$. 
Then, even though $W^1$ and $W^2$ do not annihilate the state, we get
$W^\mu W_\mu(={1 \over 2}(W^+W^-+ W^- W^+) - (W^1)^2 - (W^2)^2)  \mid k \lambda \rangle =0$ 
as it should be for a massless particle.

We have the helicity operator, 
just as in the case of massive particle,  given in Eq. (\ref{j3}).  
In analogy with 
the transverse spin for massive particles, we {\it define}
 the transverse spin 
operators for massless particles as
\begin{eqnarray}
{\cal J}^i = W^i - P^i {\cal J}^3. ~~(i=1,2)
\end{eqnarray}
They do satisfy $
{\cal J}^i \mid k, \lambda \rangle =0$, 
${\cal J}^3 \mid k, \lambda \rangle = \lambda \mid k, \lambda \rangle.$
The operators ${\cal J}^i$ and ${\cal J}^3$ obey the $E_2$-like 
algebra (Eq. (\ref{eq1})). Thus we have demonstrated that in light front
field theory, thanks to kinematic boost symmetry, it is possible to
construct spin operators for both massive and massless particles in arbitrary
reference frame.

Next, to explore the canonical spin operators in light front QCD, 
we first need to
construct the Poincare generators $P^+$, $P^i$, $P^-$, 
$K^3$, $E^i$, $J^3$ and
$F^i$ in the theory. The explicit form of the operator $J^3$ is given
Ref. \cite{hk}. Here we derive the expressions for the interaction dependent
transverse rotation operators in light front QCD. We set $A^+=0$ 
and eliminate the dependent variables $\psi^-$ and $A^-$ using the equations
of constraint. In this paper we restrict ourselves to the topologically 
trivial sector
of the theory and set the boundary condition $A^i(x^-, x^i) \rightarrow 0 $
as $ x^{-,i} \rightarrow \infty$. This completely fixes the gauge and put
all surface terms to zero.

The transverse rotation operator 
\begin{eqnarray}
F^i = {1 \over 2} \int dx^- d^2 x^\perp \Big [ x^- \Theta^{+i} - x^i
\Theta^{+-} \Big ].
\end{eqnarray}
The symmetric, gauge invariant energy momentum tensor 
\begin{eqnarray}
\Theta^{\mu \nu} &&= { 1 \over 2} {\overline \psi} \Big [ 
  \gamma^\mu i D^\nu + \gamma^\nu i D^\mu \Big ] \psi - F^{\mu \lambda a}
F^{\nu a}_{\, \, \lambda}  \\
&&~  - g^{\mu \nu} \Big [ - { 1 \over 4} (F_{\lambda \sigma a})^2 +
{\overline \psi} ( \gamma^\lambda i D_\lambda - m) \psi \Big ],
\end{eqnarray}
where 
\begin{eqnarray}
i D^\mu &&= {1 \over 2} \st{\lh}{i\pp^\mu} + g A^\mu,  \\
F^{\mu \lambda a} && = \partial^\mu A^{\lambda a} - \partial^\lambda 
A^{\mu a} + g f^{abc} A^{\mu b} A^{\lambda c},  
\end{eqnarray}
In the gauge $A^+=0$, 
using the equations of constraint
\begin{eqnarray}
i \partial^+ \psi^- && = \big [ \alpha^\perp \cdot ( i \partial^\perp + g
A^\perp) + \gamma^0 m \big ] \psi^+,   \\
{ 1 \over 2} \partial^+ A^{-a} &&= \partial^i A^{ia} + g f^{abc} { 1 \over
\partial^+}(A^{ib} \partial^+A^{ic})
 + 2 g { 1 \over \partial^+} \Big (
\xi^\dagger T^a \xi \Big ), 
\end{eqnarray}
and the equation of motion
\begin{eqnarray}
i \partial^- \psi^+ &&= -g A^- \psi^+ + \big [ \alpha^\perp \cdot  (i
\partial^\perp + g A^\perp) + \gamma^0 m \big]
{ 1 \over i \partial^+}
\big [ \alpha^\perp \cdot  (i
\partial^\perp + g A^\perp) + \gamma^0 m \big]   \psi^+,  
\end{eqnarray}
we arrive at
\begin{eqnarray}
F^2  = F^2_{I} + F^2_{II} + F^2_{III}  
\end{eqnarray}
where
\begin{eqnarray}
F^2_{I} &&= {1\over 2} \int dx^- d^2x^\p [ x^- {\cal P}^2_0 - x^2 ({\cal H}_0 +
{\cal V}) ],  \\
F^2_{II} &&= 
{1\over 2} \int dx^- d^2x^\p \Bigg [\xi^\dagger \Big [ \sigma^3 \partial^1 + i \partial^2
\Big]{ 1 \over
\partial^+} \xi   
+\Big [ { 1 \over \partial^+} (\partial^1 \xi^\dagger \sigma^3 -
i \partial^2 \xi^\dagger) \Big ] \xi \Bigg ]  \nonumber \\ 
&& ~~+ {1\over 2} \int dx^- d^2x^\p m \Bigg [ \xi^\dagger \Big [{ \sigma^1 \over i \partial^+} 
\xi\Big ] -
\Big [{ 1 \over i \partial^+} \xi^\dagger\sigma^1\Big ] \xi \Bigg ]
\nonumber  \\
&&~~+ {1\over 2} \int dx^- d^2x^\p  g \Bigg [ \xi^\dagger { 1 \over
\partial^+}[(-i \sigma^3 A^1 + A^2)\xi]  
+{ 1 \over \partial^+}
[ \xi^\dagger (i \sigma^3 A^1 + A^2)]\xi \Bigg ],  \\
F^2_{III}&&= 
- { 1 \over 2} \int dx^- d^2 x^\perp 2 (\partial^1 A^{1})A^2 
-{1\over 2} \int dx^- d^2x^\p g {4\over {\pp^+}} (\xi^\dagger T^a
\xi) A^{2a}  \nonumber \\
&&~~- {1\over 2} \int dx^- d^2x^\p g f^{abc} {2\over {\pp^+}} (
A^{ib} \pp^+ A^{ic} ) A^{2a} 
\end{eqnarray}
where $ {\cal P}^i_0$ is the free momentum density, $ {\cal H}_o$ is the
free part and ${\cal V}$ are the interaction terms in the
manifestly hermitian Hamiltonian density. $\xi$ is the non vanishing two
component of $\psi^+$.
The operators $F^2_{II}$ and $F^2_{III}$ whose integrands  do not
explicitly depend upon coordinates arise from the fermionic and bosonic
parts respectively of the gauge invariant, symmetric, energy momentum tensor
in QCD. It follows that the transverse spin
operators ${\cal J}^i$, ($i=1,2$) can also be written as the sum of three
parts, ${\cal J}^i_{I}$ whose integrand has explicit coordinate dependence, ${\cal
J}^i_{II}$ which arises from the fermionic part, and  ${\cal J}^i_{III}$ which
arises from the bosonic part of the energy momentum tensor.

Thus we have shown that even though the transverse spin operators in light
front (gauge fixed) QCD cannot be written as a sum of orbital and spin 
contributions, one
can decompose them into three distinct contributions. This is to be
contrasted with the case of helicity which can be written as a sum of
orbital and spin parts. We emphasize that our analysis is done in an
arbitrary reference frame. It is interesting to contrast our work with Ref.
\cite{jiprl} where a gauge invariant decomposition of the nucleon spin is
performed. The analysis of Ref. \cite{jiprl} is valid only in the rest frame
of the hadron and further no distinction is made between helicity and transverse
spin.

In the rest of this
paper we establish the physical relevance of this decomposition by   
exploring the connection between hadron expectation values of the
transverse spin operators and the 
quark and gluon distribution functions that
appear in transversely polarized deep inelastic scattering.

It is known that the transverse polarized distribution function in deep 
inelastic scattering is
given by (we have taken transverse polarization along the $x$-axis)
\begin{eqnarray}
g_T(x) &&= {1\over 8\pi M} \int d\eta 
		e^{-i\eta x}\times 
  \langle PS^1|\overline{
		\psi}(\eta) \Big(\gamma^1 -{P^1\over P^+}
		\gamma^+ \Big)\gamma_5 \psi(0) +~ h.c. |PS^1 \rangle 
		\, ,
\end{eqnarray}
where $P^\mu$ and 
$S^\mu$ are the four momentum and the polarization vector of the target. 
Using the constraint equation for $\psi^-$, we arrive at
\begin{eqnarray}
\int_{-\infty}^{+ \infty} dx g_T(x) && = 
\int_{-\infty}^{+ \infty} dx (g_{T(I)}(x) + g_{T(II)}(x)) 
\end{eqnarray}
\begin{eqnarray}
\int_{-\infty}^{+ \infty} dx g_{T(I)}(x) && = 
{ 1 \over 2 M} \langle P S^1 \mid
\Bigg [ \xi^\dagger \Big [ \sigma^3 \partial^1 + i \partial^2 \Big ] { 1 \over
\partial^+} \xi +
{\partial^1 \over \partial^+}(\xi^\dagger) \sigma^3 \xi - i
{\partial^2 \over \partial^+} (\xi^\dagger) \xi  \\
&&~~ + m \xi^\dagger \sigma^1 { 1 \over i \partial^+}(\xi) - m { 1 \over
i \partial^+} (\xi^\dagger) \sigma^1 \xi  \\
&& ~~+g
\Big 
[ \xi^\dagger { 1 \over
\partial^+}[(-i \sigma^3 A^1 + A^2)\xi]  
+ { 1 \over \partial^+}
[ \xi^\dagger (i \sigma^3 A^1 + A^2)]\xi \Big ]
 \Bigg] \mid P S^1 \rangle .
\end{eqnarray}
Thus the integral of $g_{T(I)}(x)$ is directly proportional to the nucleon
expectation value of $F^2_{II}$. Both $g_{T(I)}$ and $F^2_{II}$ depend on
the center of mass motion whereas both $g_T$ and ${\cal J}^i$ are
independent of the center of mass motion. The removal of the center of mass
motion from $g_{T(I)}$ is achieved by $g_{T(II)}$. We have, 
\begin{eqnarray}
\int_{-\infty}^{+ \infty} dx g_{T(II)}(x)=   
{ 1 \over M} { P^1 \over P^+}\langle P S^1 \mid\xi^\dagger 
\sigma^3 \xi \mid P S^1 \rangle. 
\end{eqnarray} 
The integral of $g_{T(II)}(x)$ is directly proportional to the nucleon
expectation value of the quark intrinsic helicity operator
\begin{eqnarray}
J^3_{q(i)} = { 1 \over 2} \int dx^- d^2 x^\perp \xi^\dagger \sigma^3 \xi.
 \end{eqnarray}

Consider the polarized gluon distribution function that appears in
transversely polarized scattering (see Ref. \cite{Ji}) 
\begin{eqnarray}
G_T(x) = { 1 \over 8 \pi x (P^+)^2} { 1 \over ({S^i}^2)} i \epsilon^{\mu \nu
\alpha \beta} S_\alpha P_\beta 
\int d \eta e^{- i \eta x  }
\langle P S^\perp \mid F^{+a}_{~~ \mu} (\eta) F^{+a}_{~~\nu}(0) \mid
P S^\perp \rangle.  
\end{eqnarray}
For a transversely polarized nucleon, $ S^+=0$. Further $F^+_{~~ -}=0$.
Since for $ \alpha, \mu , \nu =- $, the contribution is automatically zero, 
$ \beta = -$. 
Further, let us pick, without loss of generality, the transverse
polarization along the $ x $ axis. Then, $S^2=0$, $S^1=M$, $ S^-=2 {P^1 \over
P^+}M$.
Thus $\alpha$ is
forced to be $1$ or $+$. 
Then 
\begin{eqnarray}
G_T(x) = G_{T(I)}(x) + G_{T(II)}(x) 
\end{eqnarray}
where
\begin{eqnarray}
G_{T(I)}(x)  = { i \over 8 \pi x MP^+}  
\int d \eta e^{-i \eta x} 
\langle P S^1 \mid  F^{+a}_{~~ 2} (\eta) F^{+a}_{~~ +}(0) - 
F^{+a}_{~~
+}(\eta) F^{+a}_{~~ 2}(0)  \mid P S^1 \rangle,  
\end{eqnarray}
and
\begin{eqnarray}
G_{T(II)}(x)  = - 
{ i \over 16 \pi x P^+M^2} S^-  \int d \eta 
e^{-i \eta x} 
\langle P S^1 \mid F^{+a}_{~~1}(\eta) F^{+a}_{~~2}(0) -
F^{+a}_{~~2}(\eta) F^{+a}_{~~1}(0) \mid P S^1 \rangle.  
\end{eqnarray}
We get,
\begin{eqnarray}
\int_{ - \infty }^{+ \infty} dx G_{T(I)}(x)= 
 { 1 \over 2 M} \langle P S^1
\mid
A^{2a}(0) { 1 \over 2} \partial^+ A^{-a}(0) \mid P S^1 \rangle. 
\end{eqnarray} 
 From the constraint equation,
we explicitly see that the integral of
$G_{T(I)}$ is proportional to the nucleon expectation value of $F^2_{III}$.

We also have,
\begin{eqnarray}
\int_{- \infty}^{\infty} dx G_{T(II)}(x) =
 { 1 \over 4 M}{P^1 \over P^+} 
\langle P S^1 \mid \left 
(A^a_1 \partial^+ A_2^a - A^a_2 \partial^+ A_1^a \right )
\mid P S^1 \rangle . 
\end{eqnarray}
Thus the integral of $G_{T(II)}(x)$ is proportional to the 
nucleon expectation value of the
gluon intrinsic helicity operator 
\begin{eqnarray}
J^3_{g(i)} = { 1 \over 2} \int dx^- d^2 x^\perp \left [ 
A^a_1 \partial^+ A^a_2 - A^a_2 \partial^+ A^a_1 \right ]. 
\end{eqnarray}
Thus, provided the interchange of the order of
integrations is legal, we have shown that a direct relation exists 
between the
coordinate independent part of ${\cal J}^i$ which arises from the gauge
invariant fermionic and gluonic parts of the symmetric 
energy momentum tensor and the
integrals of the quark and gluon distribution functions $g_T$ and $G_T$ that
appear in polarized deep inelastic scattering.

Since transverse spin is responsible for the helicity flip of the nucleon in
light front theory, we now have identified the complete set of
operators responsible for the helicity flip of the nucleon. 
After the experimental discovery of the so-called spin crisis, the question
of the sharing of nucleon helicity among its constituents has become an
active research area. On the theoretical side, the first step involves the
identification of the complete set of operators contributing to nucleon
helicity. In this work, we have made this identification in 
the case of transverse spin. We have explicitly shown that the operators
involved in the case of the helicity and transverse spin are very
different. Because of their interaction dependence, operators contributing
to transverse spin are more interesting from the theoretical point of view
since they provide valuable information on the non-perturbative 
structure of the hadron.

With the construction,  presented in this work, of the 
transverse spin operators that commute with the mass operator, it is now
possible to explore explicitly the questions of rotational symmetry and Lorentz
invariance in light front QCD bound state calculations. 
An important issue in the case of transverse spin operators concerns
renormalization. Since they are interaction dependent, they will acquire
divergences in perturbation theory just like the Hamiltonian. It is of
interest to find the physical meaning of these divergences and their
renormalization. Of the coordinate dependent and coordinate independent
parts in the transverse spin density, the renormalization of only the
coordinate independent parts have been discussed in the literature. It is
worthwhile to recall that even in the case of ${\cal J}^3$, radiative
corrections to the coordinate dependent (orbital) parts have been studied 
only very recently. 

We plan to address the aforementioned issues 
in the future by computing the expectation value of the transverse spin 
operators in a dressed quark state. Another important problem is the
nonperturbative evaluation\cite{bk} of the matrix elements of transverse spin
operators in light front QCD.     

In summary, we have demonstrated that in light front field theory,
thanks to kinematic boost symmetry, it is possible to construct spin
operators for both massive and massless particles in arbitrary reference
frame. 
In the gauge fixed theory of light front QCD, 
we found that the transverse rotation operators
can be decomposed as the sum of three distinct terms: $F^i_{I}$ which has
explicit coordinate dependence in its integrand, and $F^i_{II}$ and $F^i_{III}$ which have no
explicit coordinate dependence. Further, $F^i_{II}$ and $F^i_{III}$ arise
from the fermionic and bosonic parts of the energy momentum tensor. 
Since transverse rotation operators are directly related to transverse spin
operators,
we 
have further shown that the expectation values of ${\cal J}^i_{II}$ and 
${\cal J}^i_{III}$ 
in a transversely polarized nucleon
are
directly related to the integral of the polarized quark and gluon
distribution
functions $g_T$ and $G_{T}$ respectively that appear in 
transversely polarized deep inelastic scattering. Finally, we have discussed
the theoretical and phenomenological consequences of our results.

We acknowledge helpful conversations with Rajen Kundu, Samir
Mallik, Partha Mitra, Jianwei Qiu and
James P. Vary. We also thank James P. Vary for a careful reading of the
manuscript. RR gratefully acknowledges the financial assistance of the
Council of Scientific and Industrial Research (CSIR), India.

\end{document}